\documentclass[12pt,preprint2]{aastex}

\shorttitle{Eccentric evolution of SMBH binaries}

\shortauthors{Iwasawa et al. }

\begin{document}

\title{Eccentric evolution of supermassive black hole binaries}

\author{Masaki Iwasawa\altaffilmark{1,2} ,
Sangyong An\altaffilmark{3} ,
Tatsushi Matsubayashi\altaffilmark{4} , \\
Yoko Funato\altaffilmark{5} ,
Junichiro Makino\altaffilmark{2}}

\email{iwasawa@strw.leidenuniv.nl}

\altaffiltext{1}{Leiden Observatory, Leiden University,
NL-2300 RA Leiden, The Netherlands}
\altaffiltext{2}{Division of Theoretical Astronomy, National
Astronomical Observatory, 2-2-1 Osawa, Mitaka, Tokyo 181-8588, Japan}
\altaffiltext{3}{LCD Business, Samsung Electronics, 
Myungamri 200, Tangjeongmyeon, Asan city, Chungnam, South Korea, 336-841}
\altaffiltext{4}{NTT Communication Science Laboratories, NTT Corporation, 
2-4 Hikaridai, Seika-cho, Soraku-gun, Kyoto, 619-0237, Japan}
\altaffiltext{5}{Department of General System Studies, University of Tokyo,
3-8-1 at Komaba, Komaba, Meguro-ku, Tokyo 153-8902, Japan}

\begin{abstract}
In recent numerical simulations \citep{matsubayashi07,lockmann08}, it
has been found that the eccentricity of supermassive black hole(SMBH)
- intermediate black hole(IMBH) binaries grows toward unity through
interactions with stellar background. This increase of eccentricity
reduces the merging timescale of the binary through the gravitational
radiation to the value well below the Hubble Time.
It also gives the theoretical explanation of the existence of eccentric
binary such as that in OJ287 \citep{lehto96, valtonen08}.
In self-consistent N-body simulations,
this increase of eccentricity is always observed.
On the other hand, the result of scattering
experiment between SMBH binaries and field stars \citep{quinlan96}
indicated no increase of eccentricity. This discrepancy leaves 
the high eccentricity of the SMBH binaries in $N$-body simulations unexplained.
Here we present a stellar-dynamical mechanism that
drives the increase of the eccentricity of an SMBH binary with large mass ratio.  
There are two key processes involved. The first one is the Kozai mechanism
under non-axisymmetric potential, which effectively randomizes the
angular momenta of surrounding stars.  The other is the selective
ejection of stars with prograde orbits.  Through these two mechanisms,
field stars extract the orbital angular momentum of the SMBH binary.
Our proposed mechanism causes the increase in the eccentricity of most
of SMBH binaries, resulting in the rapid merger through gravitational wave
radiation.
Our result has given a definite solution to  the ``last-parsec problem''
\end{abstract}


\section{Introduction}

Most of galaxies contain SMBHs at their centers. 
If a galaxy collides with another galaxy, the merging remnant hosts two SMBHs and they form a binary system.
\citet{begelman80} discussed the possibility of the formation of SMBH binaries and their evolutions.
They concluded that the evolution of the semi-major axis stalls after they ejected out all the
nearby stars (loss cone depletion) and the binary is
unlikely to merge within the Hubble time, as far as we consider the
stellar dynamical effects and the gravitational wave radiation only.
This problem is called as "last parsec problem" 
since the typical separation of a stalled binary is around one parsec.
Many researchers studied this problem and 
confirmed that the separation of the binary stall \citep{makino04,berczik05}.
Most of previous works on the evolution of the SMBH binary focused on
the evolution of the semi-major axis.
Recently, \citet{matsubayashi07} performed $N$-body
simulations of SMBH-IMBH binaries with the mass ratio of 1000:1, and
found that the eccentricity of the binary grows toward unity after the
loss cone depletion. Because of this increase of eccentricity, the
merging timescale through the gravitational radiation goes below
$10^7$ years even if the loss cone is depleted. 
Similar result was obtained by \citet{lockmann08}, and it has been reported that under certain
circumstance SMBH binaries of mass ratios close to 1:1 show similar
increase of the eccentricity \citep{aarseth08, berentzen09}.

However, this increase of the eccentricity is difficult to understand,
because the measured differential cross sections of interaction
between an SMBH binary and field stars indicated that eccentricity of a
SMBH binary would not increase through interaction with 
field stars \citep{quinlan96}.
Recently, \citet{sesana08} pointed out 
that the bound stars around SMBH binary 
play important role in the evolution of the eccentricity 
but did not give the explanation about the mechanism of the eccentricity growth.

If the increase of eccentricity actually occurs in
real SMBH binaries, it naturally explains the orbit of the binary in the
quasar OJ287 \citep{lehto96, valtonen08}.  The central SMBH of the
quasar OJ287 has the mass of $1.8 \times 10^{10} M_{\odot}$, and the
secondary black hole has the mass $\sim 10^{8} M_{\odot}$, orbital
period of 12 yrs, and eccentricity of 0.66.
Currently, this SMBH binary is in the process of circularization due
to the gravitational wave radiation. Thus, one would expect that
OJ287 had much larger semi-major axis and eccentricity
very close to unity in the past.

In this letter, we describe the mechanism which drives the increase of the eccentricity.
In section 2, we describe the numerical experiment we performed.
In section 3, we present the interpretation of the numerical result.
We give the discussion and summary in section 4.

\section{Numerical Simulation}

In order to clarify the mechanism which drives the growth of the eccentricity of
unequal-mass SMBH binaries, we performed $N$-body simulations and investigated
the change in orbital property of field stars due to interaction with 
the SMBH binary.  Our initial model is similar to that of \citet{matsubayashi07}.
Galaxy density profile is given by
\small{
  \begin{equation}
    \rho(r) = \frac{\eta}{4\pi} \frac{M_{\eta}r_0^2}{r^{3-\eta}(r_0^2+r^2)^{\eta /2 +1}} ,
  \end{equation}
}
where $r_0$ is the scale length and $M_{\eta}$ is the total mass of
field particles, and we set $r_0=60 {\rm pc}$ and $M_{\eta}=1.25 \times 10^{9} {\rm M_{\odot}}$.
From observations, the slope of central density of a giant elliptical 
galaxy is shallower than $-1$ \citep{lauer95, faber97, kormendy09}.
This feature is well understood by numerical simulations and theory \citep{ebisuzaki91, nakano99, nakano992}.
In our simulation, the density is $\rho \propto r^{-0.75}$ cusp.
The velocity profile is isotropic and consistent 
with the density profile.
The number of field stars $N$ are 16384 and 32768.
We performed five runs for each $N$.
The heavier SMBH particle whose mass is $M_p = 10^{10} {\rm M_{\odot}}$ 
is placed at the center of the galaxy, and the lighter one whose mass is
$M_s = 10^{8} {\rm M_{\odot}}$ is placed in an eccentric orbit 
with eccentricity $e$ of 0.2 and semi-major axis $a$ of 20pc.
At this radius the enclosed stellar mass is
comparable to that of the secondary mass
and the stall of the evolution of semi-major axis takes place quickly.
The mass of field particles $3.8 \times 10^4 {\rm M_{\odot}}$ in the case of $N=32768$.
Our simulation roughly mimics the OJ287 system.

The time integration scheme we used is the fourth-order Hermite scheme \citep{makino92}
for interactions between the field-field particles and the secondary SMBH - field particles.
The forces from the primary SMBH are integrated with the sixth-order Hermite scheme \citep{nitadori08} 
to reduce the integration error.
To save the calculation time,
we used GRAPE-DR \citep{makino07} for calculating interactions between the field-field particles.

Figure 1 shows the evolution of the semi-major axis $a$, the
eccentricity $e$ and the merging timescale due to the gravitational
wave radiation $T_{GW}$ of a SMBH binary using \citet{peters64} formula, for an $N=32768$ run.
The evolution of the
semi-major axis slows down and deviates from the theoretical
prediction. At the same time, the eccentricity starts to grow, and
reaches $\sim 0.99$ by $T=260$ Myrs. 
This increase of eccentricity
causes the drastic decrease of the merger timescale through gravitational wave radiation. 
In this case, at $T=180$ Myrs, the merger timescale is 10 Gyrs, 
but by $T=250$ Myrs the timescale shrinks to 10 Myrs.
We did not include the effect of the gravitational wave
radiation, but if we did, the SMBH binary would have merged very quickly.

Figure 2 shows the evolutions of the eccentricity for all runs.
The evolutions in all runs inferred from Figure 2 are basically similar.
The variation of the results of $N=32768$ runs is smaller than that of $N=16384$ runs.
These appear to be no statistically significant difference between the results for $N=16384$ and $32768$.

\section{Why does the eccentricity grow?}

We have carefully analyzed the behavior of the stars
around the SMBH binary in our numerical simulation, and found that the
following three-stage mechanism drives the increase in the
eccentricity of the secondary.

\begin{enumerate}

\item The presence of the secondary gives small, non-axisymmetric
  perturbation to the orbits of the field stars, which would be purely
  Keplerian otherwise.  Because of this perturbation the orbital
  motions of the stars become chaotic, and their angular momenta
  change chaotically.

\item Since the orbital angular momenta of stars are not conserved,
  they can experience close encounters with the secondary, some of
  which result in the ejection of stars. When a star is on a prograde
  orbit relative to the orbit of the secondary, it has a much larger
  chance of being ejected than when it is on a retrograde orbit.

\item Since the secondary tends to scatter stars on prograde orbits,
  there will be  more stars on retrograde orbits
  than on prograde orbits around the binary.  However, the
  chaotic changes of angular momenta of stars reduce the difference between
  the number of stars on prograde and retrograde orbits. This means the
  angular momentum of the secondary is transferred to field stars and
  the eccentricity of the secondary increases.
\end{enumerate}

Note that in this process the transfer of angular momentum occurs
between the binary and field stars, while field stars are bound to the
system, through secular perturbation. This secular perturbation is
neglected in previous theoretical models on the behavior of SMBH
binaries \citep{begelman80} or models based on scattering
experiment \citep{quinlan96}, 
and can give the reasonable interpretation of the results of \citet{sesana08}.

Figure 3 shows the time evolution of the specific angular
momentum of a star that comes close to the secondary SMBH.  The orbital
plane of the secondary SMBH is the $x-y$ plane.  One can see that none
of the three components of the angular momentum are conserved and the
changes are complicated.  These changes are due to the secular
perturbation of the secondary SMBH. If the secondary SMBH had a
circular orbit, the secular perturbation would not have any
non-axisymmetric term, and therefore $L_z$ would be
conserved \citep{kozai62}. However, since the secondary SMBH has
non-zero eccentricity, the perturbation has non-axisymmetric term,
resulting in a triaxial potential field.  Thus, the orbits of many
stars become chaotic \citep{ford00,levin05}.

This chaotic change of the orbits of field stars means that they
approach the secondary SMBH when their total angular momentum becomes
small enough. One can assume that stars on both prograde and
retrograde orbits will approach the secondary SMBH in roughly equal
chances. The cross section of the large change of the binding energy,
which would result in the ejection of a star, is larger for prograde
orbits, because the typical relative velocity is smaller for prograde
orbits and the effect of gravitational focusing is larger.

Figure 4 shows the cumulative mass of ejected stars in units of the mass 
of secondary SMBH. We can see that prograde stars are more likely to be 
ejected.  
In this calculation, the difference between the cumulative mass of
ejected stars on prograde and retrograde orbits was comparable to the
mass of the secondary SMBH.  For $T>200 {\rm Myrs}$, the ejection rates
of prograde and retrograde stars are rather similar, but that is
simply because the eccentricity of the secondary SMBH has become large
enough and therefore the relative difference in the cross section
terms has become small.

Figure 5 shows the evolution of the average value of $L_z$ of stars
which are bound to the primary SMBH at $100 {\rm Myrs}$ 
and are ejected by $200 {\rm Myrs}$.
The net change of $L_z$ of stars which were initially in prograde orbit is 
small, while that of stars in retrograde orbits is large.
Thus the primary mechanism of angular momentum transfer is the transfer
of angular momentum to retrograde stars through secular perturbation 
and removal of them through ejection.

To summarize, we found a quite efficient mechanism that causes the
increase of the eccentricity of unequal-mass SMBH binaries.  It is the
combination of the non-axisymmetric potential field of the secondary
SMBH, which randomizes the angular momentum of field stars, and the
selective ejection of stars with prograde orbits through close encounters
with the secondary SMBH. 
Since either effects depends on masses of field particles,
our proposed mechanism does not depend on $N$.

This mechanism is effective at least for SMBH
binaries with fairly large mass ratios (larger than 10:1).
There is no other known mechanism that can make the eccentricity of a
SMBH binary close to unity. Both the gravitational wave radiation
and gas dynamical effect reduce the eccentricity, and interaction
with field stars, if we neglect the mechanism we described in this
paper, would not change the eccentricity \citep{quinlan96}.  If the
parent galaxy is globally nonspherical, the SMBH binary could become
eccentric because of the Kozai mechanism, but it would be hard to
reach very high eccentricity values since the $z$-component would be
conserved if the parent galaxy is axisymmetric. If the parent galaxy
is triaxial, we could expect to see a significant effect on the evolution
of the semi-major axis \citep{berczik06}, but its effect on eccentricity 
is currently unknown.  Furthermore, it is difficult to maintain 
triaxiality given the presence of the central SMBH \citep{pfenniger89,hozumi05}.

\section{Discussion and Summary}

\subsection{Equilibrium Value of Eccentricity}
If the mass ratio between the secondary SMBH and the field stars
is infinite, in other words, in the continuous limit,
we can expect that the eccentricity of the SMBH binary reaches unity.
In practice, since the mass and the number of field stars are both finite,
they give random perturbations to the angular momentum of SMBH binary.
If we assume that there is thermal equilibrium between tangential velocity 
of the secondary SMBH and random velocity of stars around the secondary, 
we can estimate the specific angular momentum $L_{S}$ and the eccentricity $e$ of the secondary SMBH as
\small{
  \begin{equation}
    L_{S} \sim \sqrt{\frac{M_{FS}}{M_{S}}}<L_{FS}>,
  \end{equation}
}
and 
\small{
  \begin{equation}
    1-e \sim \left( \frac{L_{S}}{<L_{FS}>} \right)^2 \sim \frac{M_{FS}}{M_{S}},
  \end{equation}
}
where $<L_{FS}>$ is the mean value of the specific angular momentum of the field star 
around the secondary SMBH 
and $M_{S}$ and $M_{FS}$ are masses of the secondary and the field star, respectively.

Our results in Figure 2 is consistent with these theoretical estimate.
The eccentricity of the real SMBH binary can become much close to unity,
since the mass ratio to the field star is much larger.

\subsection{Merging Timescale}

The timescale of the evolution of the eccentricity,
$T_e = (1-e) / |\dot e|$, due to our proposed mechanism is much
smaller than the Hubble time.  

The timescale of the change in the
angular momentum of the field stars is proportional to $M_p/M_s$,
where $M_p$ and $M_s$ are the masses of the primary and secondary SMBHs,
respectively. The rate at which angular momentum is removed 
through ejection of stars is $\propto (M_s/M_p)^2$, since it is proportional to
the interaction cross section. Therefore, if $M_p/M_s$ is
large, the timescale of the change in the angular momentum is shorter
than that of the removal of the field stars. 
Thus $T_e$ is determined by the removal timescale.

The loss rate of the angular momentum of the secondary SMBH 
is proportional to $\rho V_s \sigma$, where $\rho$ is the density at the radius of the secondary,
$V_s$ and $\sigma$ are a Kepler velocity and cross section of the secondary.
Thus, $T_e$ is proportional to $M_s/(\rho V_s \sigma)$.
From our numerical result, we can estimate the
timescale as
\small{
  \begin{eqnarray}
    T_e &\sim & {\rm 1.4\times 10^{\rm 8}}\left(\frac{M_p}{10^{10}M_{\odot}}\right)^{3/2}
    \left(\frac{M_s}{10^8M_{\odot}}\right)^{-1} \nonumber \\
    && \left(\frac{a}{10{\rm pc}}\right)^{-3/2}
    \left(\frac{\rho}{100 M_{\odot}/{\rm pc^3}}\right)^{-1}{\rm yrs} \\ 
    &\sim & {\rm 2.0\times 10^{\rm 8}}\left(\frac{M_p}{10^{10}M_{\odot}}\right)^{3/2}
    \left(\frac{M_s}{10^8M_{\odot}}\right)^{-2} \nonumber \\
    && \left(\frac{a_{stall}}{4{\rm pc}}\right)^{3/2} {\rm yrs},
  \end{eqnarray}
}

where $a_{stall}$ is a semi-major axis of a stalled binary and 
assumed $a_{stall} \propto (M_s / \rho)^{1/3}$.

Since the majority of galaxy-galaxy merger events are minor mergers
with mass ratios of around 10:1 or more, the typical mass ratio of the
SMBH binary is also around 10:1, if we assume the linear relationship
between the SMBH mass and the mass of the stellar spheroid
\citep{Magorrian98, marconi03}.  Therefore, we can conclude that our
mechanism is effective for most of SMBH binaries, and they can merge
through interactions with field stars and gravitational wave radiation.
Our result finally has given a definite solution to  the ``last-parsec problem''
\citep{begelman80,makino04,berczik05}.

\subsection{Observation of SMBH Binary}

The mechanism we found naturally explains why the OJ287 system is eccentric.
Most likely, this mechanism resulted in the orbit of secondary with
the periastron distance similar to the present value, but with much
larger semi-major axis and eccentricity, and gravitational wave
radiation have driven the semi-major axis to its present value.

The state of a recently found candidate SMBH binary
\citep{borosonlauer2009} might also be explained by our mechanism. This
system is a binary with the mass ratio of around 30:1.  Interaction
with field stars cannot drive its semi-major axis to the value
suggested by the observed relative velocity of $3000 {\rm km~s^{-1}}$,
if we assume a circular orbit.  However, our mechanism implies that
all unequal-mass SMBH binaries with mass ratio larger than 10:1
become highly eccentric. Thus, they all
have very large relative velocities at the pericenter. Therefore the
observed relative velocity is only weakly coupled to the semi-major
axis.

Once the binary becomes highly eccentric, the gravitational wave should
be observed at each pericenter passage. The amplitude is given by
\small{
  \begin{equation}
    h \sim {\rm 10^{\rm -15}}\left(\frac{M_p}{10^{10}M_{\odot}}\right)
    \left(\frac{M_s}{10^8M_{\odot}}\right)
    \left(\frac{r_p}{1000 {\rm AU}}\right)^{-1}
    \left(\frac{R}{1{\rm Gpc}}\right)^{-1}.
  \end{equation}
}

The frequency $f$ is given by
\small{
  \begin{equation}
    f \sim {\rm 10^{\rm -7}}\left(\frac{M_p+M_s}{10^{10}M_{\odot}}\right)^{0.5}
    \left(\frac{r_p}{1000{\rm AU}}\right)^{-1.5}
         {\rm Hz},
  \end{equation}
}
where $r_p$ and $R$ are the pericenter
distance of the secondary SMBH from the primary 
and the distance of the binary from us.
Unfortunately, the frequency is too low to be observed even with
space missions like LISA. However, since the amplitude is very large,
other techniques such as Doppler tracking of interplanetary spacecraft
or pulsar timing could be used.

\subsection{Summary}
In this letter, we present a stellar-dynamical mechanism that
drives the increase of the eccentricity of an SMBH binary.  
There are two key processes involved. The first one is the Kozai mechanism
under non-axisymmetric potential, which effectively randomizes the
angular momenta of surrounding stars.  The other is the selective
ejection of stars with prograde orbits.  Through these two mechanisms,
field stars extract the orbital angular momentum of the SMBH binary.
Our proposed mechanism causes the increase of the eccentricity of most
of SMBH binaries, resulting in the rapid merger through gravitational wave radiation.

\begin{figure}
\epsscale{0.8}
\plotone{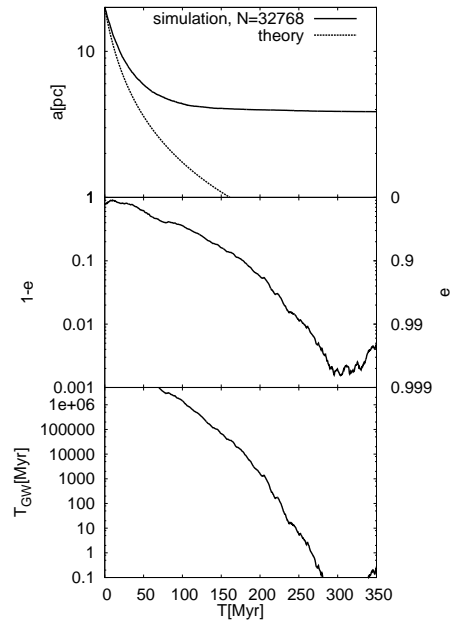}
\caption{
Evolution of semi-major axis $a$ (top panel),
eccentricity $e$ (middle panel, $1-e$ is shown) and
the merging timescale through the gravitational radiation (bottom panel) for an $N=32768$ run.
In the top panel, the thin dashed curve shows the theoretical prediction
obtained using the standard dynamical friction formula.
\label{fig1}
}
\end{figure}

\begin{figure}
\epsscale{1.0}
\plotone{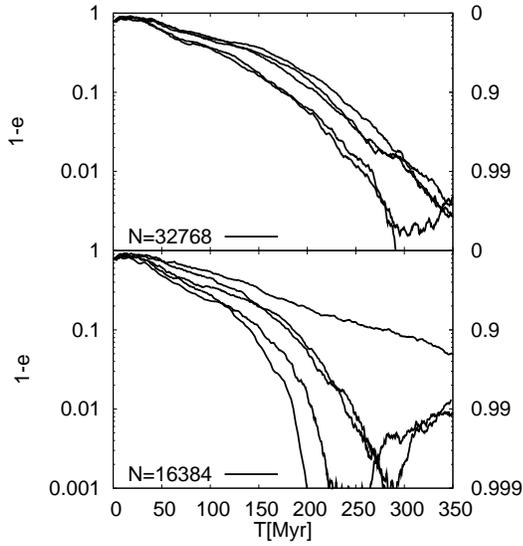}
\caption{
Evolution of eccentricity $e$ ($1-e$ is shown) for all runs.
Top and bottom panels show the results of simulations with $N$ of 32768 and 16384, respectively.
\label{fig2}
}
\end{figure}

\begin{figure}
\epsscale{0.8}
\plotone{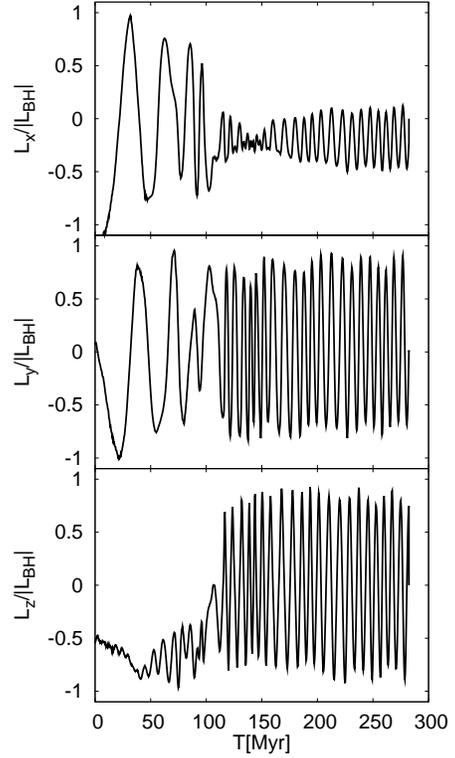}
\caption{
Time evolution
of x (top panel), y (middle panel) and z (bottom panel) components
of the specific angular momentum of a star 
(in units of specific angular momentum of the secondary SMBH at $T=0$). 
\label{fig3}
}
\end{figure}

\begin{figure}
\epsscale{1.0}
\plotone{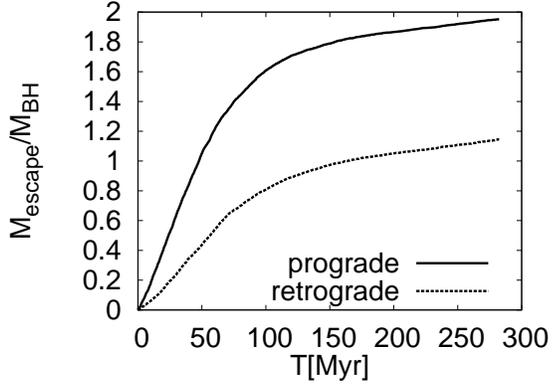}
\caption{
Cumulative mass of the ejected stars in units of the mass of the secondary SMBH.
Solid (Dashed) curve shows the number
for stars in prograde (retrograde) orbit at the time of being ejected from the
SMBH binary.
\label{fig4}
}
\end{figure}

\begin{figure}
\epsscale{1.0}
\plotone{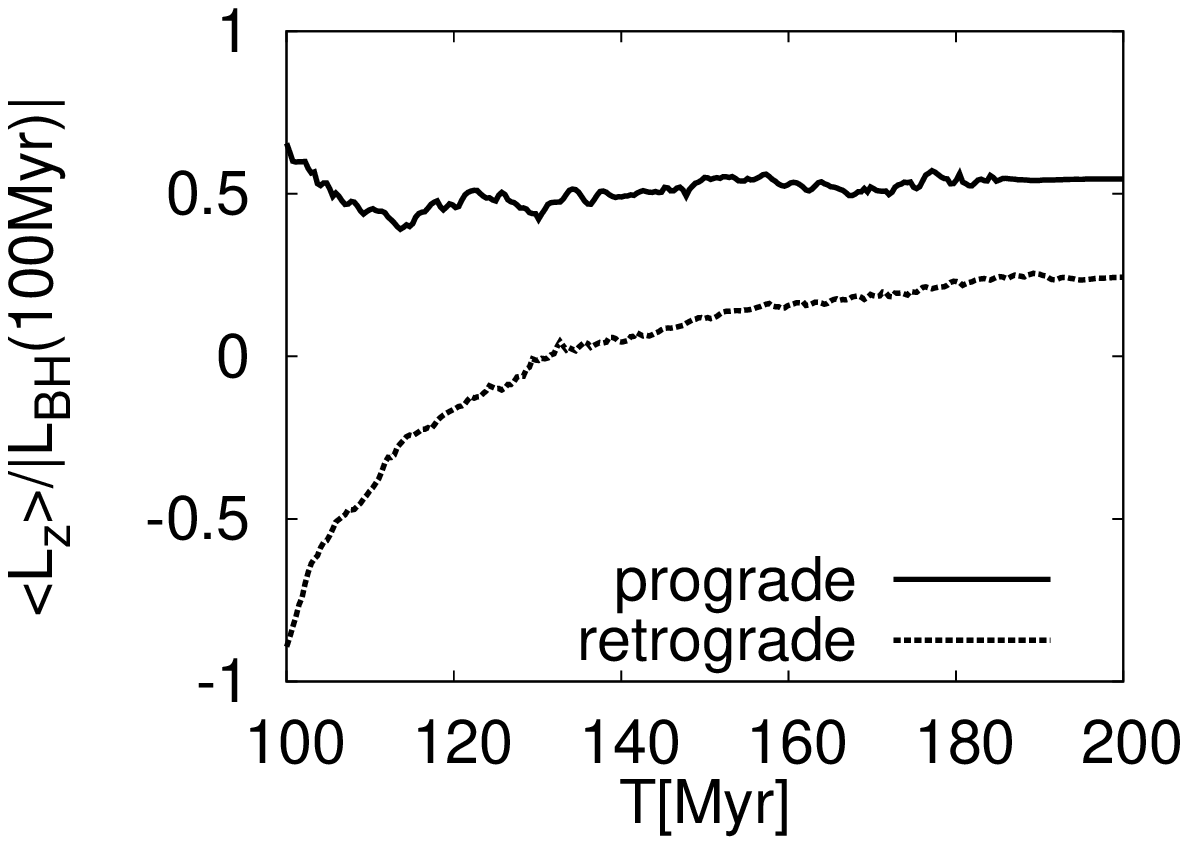}
\caption{
Evolution of the average value of $L_z$ of stars,
which are bound to the primary SMBH at $100 {\rm Myrs}$ 
and are ejected by $200 {\rm Myrs}$, 
in units of specific angular momentum of the secondary SMBH at $T=100 {\rm Myrs}$.
Solid and Dashed curves show the average value of initially 
prograde and retrograde particles, respectively.
\label{fig5}
}
\end{figure}

\acknowledgments

We thank Toshiyuki Fukushige, 
Yuichiro Sekiguchi, Ataru Tanikawa, 
Alberto Sesana, Mauri Valtonen,
Michiko Fujii, Kuniaki Koike 
and Yusuke Tsukamoto
for stimulating discussions and useful comments.

This research is partially supported by the Special Coordination 
Fund for Promoting Science and Technology (GRAPE-DR project), 
Ministry of Education, Culture, Sports, Science and Technology, Japan. 
Part of calculations were done using the GRAPE system at the
Center for Computational Astrophysics (CfCA) of 
the National Astronomical Observatory of Japan.

\end{document}